\documentclass[conference]{IEEEtran}
\IEEEoverridecommandlockouts

\usepackage{cite}
\usepackage{amsmath,amssymb,amsfonts}
\usepackage[ruled,linesnumbered]{algorithm2e}
\usepackage{graphicx}
\usepackage{textcomp}
\usepackage{xcolor}
\usepackage{nicefrac}

\title{Diffusion Models for Polarimetric Reconstruction of Circumstellar Environments in \\Correlated Speckle Noise}

\author{
  \IEEEauthorblockN{
    Quentin Villegas\IEEEauthorrefmark{1},
    Laurence Denneulin\IEEEauthorrefmark{2},
    Simon Prunet\IEEEauthorrefmark{1},
    André Ferrari\IEEEauthorrefmark{1},
    Éric Thiébaut\IEEEauthorrefmark{3},
    Maud Langlois\IEEEauthorrefmark{3},
  }
  \IEEEauthorblockA{
    \IEEEauthorrefmark{1}Université Côte d'Azur, Observatoire Côte d'Azur, CNRS, Nice, France \\
    \IEEEauthorrefmark{2}Laboratoire de Recherche de l'EPITA, EPITA, Le Kremlin-Bicêtre, France\\
    \IEEEauthorrefmark{3}Univ. Lyon 1, ENS de Lyon, CRAL (CNRS), Lyon, France
  }
}
\begin{document}

\maketitle

\begin{abstract}
High-contrast polarimetric imaging of circumstellar disks is severely limited by stellar leakage and speckle noise. Building upon the RHAPSODIE framework for polarimetric inverse problems, diffusion-based methods have shown promising results by replacing classical regularization with learned priors. However, RHAPSODIE's white noise assumption fails to capture the spatial correlation structure of atmospheric and instrumental speckles. We extend this framework to handle speckle fluctuations, which are modeled as a Gaussian field with a stationary covariance matrix with a Gaussian spectral density. To efficiently apply the dense precision matrix that includes correlated speckle and heteroskedastic photon noise, we develop a dedicated preconditioner for conjugate gradient iterations. The diffusion prior is trained exclusively on disk-only data. 
While limitations remain at very low SNRs, our method significantly outperforms Tikhonov regularization and competes favorably with a Plug-and-Play baseline in realistic regimes, successfully recovering morphological features masked by correlated noise.

\end{abstract}


\section{Introduction}

Observing circumstellar disks is fundamental to understanding planetary formation, but remains challenging due to the extreme intensity contrast with the host star, which is typically 1~000 to 10~000 times brighter. Despite advanced coronagraphic instruments such as SPHERE/IRDIS \cite{beuzit_sphere_2019}, observations remain degraded by various instrumental and atmospheric factors.

RHAPSODIE \cite{denneulin_rhapsodie_2021} brought significant improvements in circumstellar image reconstruction in polarimetry through an inverse problem approach, enabling better handling of missing data and noise. Subsequently, deep learning techniques based on unrolled neural networks have been proposed to further improve these reconstructions \cite{chappon_linear_2024}. Although effective, these methods require computing gradients through all iterations during training, resulting in a large memory footprint, and must be retrained for each new instrumental configuration. 
Plug-and-Play (PnP) methods \cite{venkatakrishnan_plug-and-play_2013, chan_plug-and-play_2017} also offer an interesting alternative, although they require careful hyperparameter tuning \cite{wei_tuning-free_2020}.

In this work, we propose an approach based on diffusion models \cite{ho_denoising_2020} for the reconstruction of circumstellar disks from polarimetric observations. Rather than relying on classical regularization, we exploit the capacity of denoising diffusion probabilistic models to learn the statistical distribution of disk images, and adapt the Score-ALD (Annealed Langevin Dynamics) algorithm \cite{jalal_robust_2021} for conditional sampling from degraded observations. The key contribution is to extend this framework to correlated speckle noise, and to develop a Fourier-domain preconditioner that makes Score-ALD practical at realistic image scales.

Moreover, RHAPSODIE assumes spatially uncorrelated (but not stationary) noise, modeled by a diagonal precision matrix. This assumption fails to capture the spatial correlation structure of atmospheric and instrumental speckles, which are inherently correlated interference patterns at the diffraction limit. At moderate to high speckle levels originating from starlight, this model mismatch leads to structured artifacts that can be mistaken for genuine disk features. To address this limitation, we extend the forward model to explicitly separate the disk signal from stellar speckle contamination, requiring as additional input an estimate of the mean speckle pattern. Speckle fluctuations around this mean are modeled as a Gaussian random field with a parameterized power spectral density. Although this Gaussian model allows unphysical negative fluctuations, it remains a tractable approximation that effectively captures the spatial correlation structure of speckles. The resulting hybrid covariance structure requires dedicated numerical solvers, for which we develop an efficient preconditioned conjugate gradient method.

The major advantage of our approach is its ability to generate samples from the posterior distribution. By combining conditional sampling with an extended observation model that accounts for both stellar leakage and correlated speckle noise, the proposed method effectively reconstructs circumstellar structures even from heavily corrupted data. Numerical results demonstrate a significant improvement in the recovery of disk morphological details, opening new perspectives for the detailed study of circumstellar environments.

The remainder of this paper is organized as follows. Section~II presents the extended forward model with correlated speckle noise. Section~III gives an overview of the proposed method and introduces the reconstruction algorithm. Section~IV derives the likelihood gradient for the Bayesian inverse problem. Section~V develops an efficient Fourier-domain preconditioned conjugate gradient solver for the correlated noise linear systems. Section~VI describes the diffusion prior and its training. Section~VII presents experimental results, and Section~VIII concludes.

\begin{table}[t]
\centering
\caption{Summary of key notation}
\begin{tabular}{cl}
\hline
Symbol & Description \\
\hline
$x \in \mathbb{R}^{2N^2}$      & Unknown disk image $(I^u, I^p)$ \\
$y_k \in \mathbb{R}^{2N^2}$    & Observation frame $k$ \\
$\mathsf{A}_k = H_k K S_k$     & Forward operator (frame $k$) \\
$\mathsf{B}_k = H_k S_k$       & Speckle operator (no PSF) \\
$\boldsymbol{\Sigma}_n$         & Diagonal photon/detector noise covariance \\
$C_{\delta s}$                  & Speckle fluctuation covariance (Fourier-diag.) \\
$C_{\varepsilon,k}$             & Total noise covariance (frame $k$) \\
$W_k = C_{\varepsilon,k}^{-1}$ & Precision matrix (frame $k$) \\
$\bar{s}$                       & Mean speckle pattern \\
$P_0,\, \sigma_\text{corr}$    & PSD parameters (variance, correlation length) \\
\hline
\vspace{-2em}
\end{tabular}
\label{tab:notation}
\end{table}

\section{Forward Model and Correlated Noise}

\subsection{Polarimetric Imaging and RHAPSODIE Framework}

Fundamentally, the reconstruction problem reduces to a linear inverse problem:
given $K$ observation vectors $y_k \in \mathbb{R}^{2N^2}$, recover the unknown image $x \in \mathbb{R}^{2N^2}$ such that:
\begin{equation}
    y_k = \mathsf{A}_k x + n_k, \quad k = 1, \dots, K
\end{equation}

where $\mathsf{A}_k$ is a known linear operator encoding the instrument physics, and $n_k$ is additive noise. The specific structure of $\mathsf{A}_k$ and the noise model $n_k$ are detailed below; readers unfamiliar with polarimetric imaging may focus on this abstract formulation and treat $\mathsf{A}_k$ as a generic measurement matrix.

SPHERE-IRDIS \cite{beuzit_sphere_2019} acquires a sequence of $K$ dual-channel observations by rotating a half-wave plate (HWP) at angles $\alpha_k \in \{0^\circ, 22.5^\circ, 45^\circ, 67.5^\circ\}$. Each frame $k$ yields left ($\xi_k$) and right ($\eta_k$) channel intensities through polarimetric analyzers at $\psi_j \in \{0^\circ, 90^\circ\}$. Following RHAPSODIE \cite{denneulin_rhapsodie_2021}, the observation model for frame $k$ is:
\begin{equation}
y_k = \mathsf{A}_k x + n_k
\label{eq:rhapsodie_basic}
\end{equation}
where $y_k = (\xi_k; \eta_k) \in \mathbb{R}^{2N^2}$ concatenates both detector images with $N \times N$ pixels, and $n_k \sim \mathcal{N}(0, \boldsymbol{\Sigma}_n)$ models combined detector and photon noise, with $\boldsymbol{\Sigma}_n$ diagonal (spatially uncorrelated). The Stokes parameters relate to our variables through $I = I^u+I^p$, $Q = I^p \cos(2\theta)$, and $U = I^p \sin(2\theta)$, where $\theta$ denotes the polarization angle. Under the single-scattering assumption for a centered point source, $\theta$ is known (orthoradial polarization) \cite{engler_detection_2018}, reducing the unknowns from the full Stokes vector $(I, Q, U)$ to $x = (I^u, I^p) \in \mathbb{R}^{2N^2}$.

The forward operator $\mathsf{A}_k = H_k K S_k$ decomposes as:
\begin{itemize}
\item \textbf{Polarimetric modulation} ($S_k$): Projects $(I^u, I^p)$ onto detector channels. The operator $S_k \in \mathbb{R}^{2N^2 \times 2N^2}$ is block-diagonal, applying the same local transformation $S_k^{\text{loc}} \in \mathbb{R}^{2 \times 2}$ at each pixel:
\vspace{-0.5em}
\begin{equation}
\begin{bmatrix} \tilde{\xi}_k \\ \tilde{\eta}_k \end{bmatrix} = 
\underbrace{
\frac{1}{2}
\begin{bmatrix}
1 & 1 + \cos(4\alpha_k - 2\theta - 2\psi_1) \\
1 & 1 + \cos(4\alpha_k - 2\theta - 2\psi_2)
\end{bmatrix}
}_{S_k^{\text{loc}}} 
\begin{bmatrix} I^u \\ I^p \end{bmatrix}
\label{eq:modulation}
\end{equation}

with $\psi_1 = 0^\circ$ (left channel) and $\psi_2 = 90^\circ$ (right channel), corresponding to the Malus law formulation of Eq.~(8) in \cite{denneulin_rhapsodie_2021}.

\item \textbf{PSF convolution} ($K$): Toeplitz block-circulant matrix modeling instrumental blur, assumed rotationally/temporally invariant (independent of $k$).

\item \textbf{Geometric transforms} ($H_k$): Field rotation, de-rotation, and pixel alignment for both channels \cite{denneulin_rhapsodie_2021}.
\end{itemize}

Handling invalid pixels (dead, saturated, masked $\approx$10\% of data) in the presence of correlated noise is left for future work.

\subsection{Extended Model with Correlated Speckle Noise}
\label{subsec:extended-model-correlated-speckle}

The uncorrelated noise assumption in (\ref{eq:rhapsodie_basic}) fails to capture the spatial correlation of atmospheric and instrumental speckles. We model the stellar field as time-varying: $s_k = \bar{s} + \delta s_k$, where $\bar{s}$ is the mean pattern and $\delta s_k \sim \mathcal{N}(0, C_{\delta s})$ represents correlated fluctuations. Assuming speckles are unpolarized, we introduce the speckle operator $\mathsf{B}_k = H_kS_k$, identical to $\mathsf{A}_k$ but without the PSF convolution. The observation model becomes:
\begin{align}
y_k &= \mathsf{A}_k x + \mathsf{B}_k(\bar{s} + \delta s_k) + n_k \\
\label{eq:extended_model}
&= \mathsf{A}_k x + \mathsf{B}_k\bar{s} + \underbrace{\mathsf{B}_k\delta s_k + n_k}_{\varepsilon_k}
\end{align}
where the total noise $\varepsilon_k = \mathsf{B}_k\delta s_k + n_k$ has covariance:
\begin{equation}
C_{\varepsilon,k} = \mathsf{B}_kC_{\delta s}\mathsf{B}_k^T + \boldsymbol{\Sigma}_n
\label{eq:covariance}
\end{equation}

Both channels observe the same speckle field $\delta s_k \in \mathbb{R}^{N^2}$, 
so $\mathsf{B}_k C_{\delta s} \mathsf{B}_k^T \in \mathbb{R}^{2N^2 \times 2N^2}$ 
has rank $N^2$ — half the ambient dimension — making direct inversion 
of $C_{\varepsilon,k}$ intractable for typical image sizes ($2N^2 \approx 33\text{k}$); 
see Section~\ref{sec:pcg}.

\textbf{Speckle statistics.} We parameterize $C_{\delta s}$ via its power spectral density (PSD), assuming a homogeneous, isotropic Gaussian field. Using the discrete 2D Fourier transform operator $F$:
\begin{equation}
    C_{\delta s} = F^{-1}\Lambda F
\end{equation}
with $\Lambda  = \text{diag}(\lambda_j)$ and $\lambda_j = P_0 \exp(-\sigma_{\text{corr}}^2 \|f_j\|^2)$. Here, $P_0$ controls the variance, $\sigma_{\text{corr}}$ the correlation length, and $f_j$ the spatial frequency vector associated with pixel $j$. The precision matrix $W_k = C_{\varepsilon,k}^{-1}$ is no longer diagonal, and direct inversion is intractable for typical image sizes ($N \approx 128$, $M = 2N^2 \approx 33$k pixels).

\section{Overview of the Proposed Method}

The proposed reconstruction pipeline combines an extended observation 
model accounting for correlated speckle noise with a diffusion-based 
prior on disk morphology. The two core technical contributions are: 
(i) a Fourier-domain preconditioner that enables efficient resolution 
of the correlated noise linear system, and (ii) the adaptation of 
Score-ALD \cite{jalal_robust_2021} to this extended observation model 
with non-diagonal precision matrices.

The full procedure is summarized in Algorithm~\ref{alg:score_ald_compact}. 
Starting from pure Gaussian noise $x_T \sim \mathcal{N}(0,I)$, the 
algorithm iterates over $T$ diffusion timesteps, with schedule parameters 
$T$, $\{\beta_t\}$, and $\{\bar{\alpha}_t\}$ fixed at training time. 
At each step $t$, the conditional score $\nabla_{x_t} \log p(x_t | y)$ 
is decomposed into two terms following Bayes' rule:
\begin{equation}
\nabla_{x_t} \log p(x_t|y) = \underbrace{\nabla_{x_t} \log p(x_t)}_{\text{prior score}} 
+ \underbrace{\nabla_{x_t} \log p(y|x_t)}_{\text{likelihood score}}
\end{equation}
The prior score is estimated by the trained denoising network 
$\epsilon_\theta$ (detailed in Section~\ref{sec:diffusion}). The 
likelihood score requires solving $K$ correlated linear systems 
$C_{\varepsilon,k} z_k = r_k$ at every step --- a potentially 
prohibitive cost at scale $N=128$. This is addressed by the 
Fourier-domain preconditioner derived in Section~\ref{sec:pcg}, 
which reduces the conjugate gradient iteration count by a factor 
of $5\times$. The Langevin update then reads:
\begin{equation}
x_{t-1} = x_t + \sqrt{2\beta_t}\,\zeta_t 
+ \beta_t\!\left[\frac{-\epsilon_\theta(x_t,t)}{\sqrt{1-\bar{\alpha}_t}} 
- \mu\,\mathsf{A}^T z\right]
\end{equation}
where $\zeta_t \sim \mathcal{N}(0, I)$ is the stochastic noise injection, 
and $\mu>0$ is a hyperparameter balancing the prior and likelihood scores. 
Since $\left\|\nabla_x \log p\left(y \mid x_t\right)\right\| \propto P_0^{-1/2}$ 
in the speckle-dominated regime, we empirically set $\mu \approx \sqrt{P_0}$ 
to maintain a consistent balance across speckle levels.

\begin{algorithm}[t]
\SetAlgoLined
\KwIn{Observations $\{y_k\}_{k=1}^K$, mean speckle $\bar{s}$, PSD parameters $(P_0, \sigma_{\mathrm{corr}})$, weight $\mu$}
\KwOut{Reconstructed disk $x_0$}
Initialize $x_T \sim \mathcal{N}(0, I)$\\
\For{$t = T, T-1, \ldots, 1$}{
    $\zeta_t \sim \mathcal{N}(0, I)$\\
    $s_{\mathrm{prior}} \leftarrow -\frac{\epsilon_\theta(x_t, t)}{\sqrt{1-\bar{\alpha}_t}}$\\
    $\quad r \leftarrow \mathsf{A} x_t + \mathsf{B}\bar{s} - y$\;
    $\quad z \leftarrow \mathrm{PCG}(C_\varepsilon, r, M^{-1})$ \hfill {\small // $M^{-1}$ via (\ref{eq:precond_inv})--(\ref{eq:fourier_preconditionner})}\\
    $s_{\mathrm{like}} \leftarrow -\mathsf{A}^T z$\\
    $x_{t-1} \leftarrow x_t + \sqrt{2\beta_t}\zeta_t + \beta_t\left(s_{\mathrm{prior}} + \mu\, s_{\mathrm{like}}\right)$
}
\Return $x_0$
\caption{Score-ALD with Correlated Noise}
\label{alg:score_ald_compact}
\end{algorithm}

\section{Inverse Problem and Gradient Computation}
\label{sec:gradient_comp}

We formulate the reconstruction as a Bayesian inverse problem where the posterior distribution $p(x|y) \propto p(y|x)p(x)$ combines the observations $y = (y_1, \ldots, y_K)$ with a prior $p(x)$ on the disk structures. Under the assumption of temporal independence for the speckle fluctuations $\delta s_k$, the log-likelihood is the sum of individual log-likelihoods for each frame. 

The gradient of the log-likelihood with respect to the disk image $x$ is:
\begin{equation}
\nabla_x \log p(y|x) = -\sum_{k=1}^K \mathsf{A}_k^T W_k \underbrace{(\mathsf{A}_k x + \mathsf{B}_k\bar{s} - y_k)}_{r_k} = -\mathsf{A}^T z
\label{eq:full_gradient}
\end{equation}
where $r_k$ is the residual for frame $k$, and $\mathsf{A}$ denotes the stacked operator across all $K$ frames. The vector $z$ represents the concatenated inverse covariance weighted residuals $z_k = W_k r_k$. 

In this correlated noise framework, computing the gradient requires evaluating the action of the precision matrices $W_k = C_{\varepsilon,k}^{-1}$. Since direct inversion is intractable, evaluating the gradient implies solving the linear systems:
\begin{equation}
C_{\varepsilon,k} z_k = r_k, \quad \text{for } k=1, \dots, K
\label{eq:solve_system}
\end{equation}
The efficient resolution of these systems using a specialized numerical strategy is the subject of Section~\ref{sec:pcg}.

\section{Preconditioned Conjugate Gradient Solver}
\label{sec:pcg}

\subsection{Problem and Strategy}

At each gradient evaluation, we solve $C_{\varepsilon,k} z_k = r_k$ using Conjugate Gradient, which converges in $O(\sqrt{\kappa})$ iterations where $\kappa(C)$ is the condition number of the matrix \cite{nocedal_numerical_2006}. Without preconditioning, convergence is slow—problematic since this linear system is solved thousands of times during optimization. We precondition with an approximation $M \approx C_{\varepsilon,k}$ that retains the essential structure of $C_{\varepsilon,k}$ while admitting an analytically tractable inverse $M^{-1}$, achieving better conditioning and significantly reducing the iteration count.

\subsection{Analytical Preconditioner}
\label{sec:preconditioner}

Since speckles are assumed unpolarized (as established in Section~\ref{subsec:extended-model-correlated-speckle}), the modulation matrix $S_k^{\text{loc}}$ reduces to the factor $\frac{1}{2}$ on each channel, yielding a rank-one polarimetric kernel:
\begin{equation}
S_k\, C_{\delta s}\, S_k^T = \begin{bmatrix}\nicefrac{1}{2}\\\nicefrac{1}{2}\end{bmatrix} C_{\delta s} \begin{bmatrix}\tfrac{1}{2} & \tfrac{1}{2}\end{bmatrix} = \tfrac{1}{4}\begin{bmatrix}C_{\delta s} & C_{\delta s}\\C_{\delta s} & C_{\delta s}\end{bmatrix}
\label{eq:rank1_kernel}
\end{equation}
where $C_{\delta s} = F^T \Lambda F$ is the spatial speckle covariance diagonalized in the Fourier domain with spectral coefficients $\lambda_j$. We further approximate the model under two assumptions: stationary detector noise ($\Sigma_n \approx \bar{\sigma}^2 I$) and unitary geometric transforms ($H_\xi^T H_\xi = H_\eta^T H_\eta = I$). Substituting~\eqref{eq:rank1_kernel} into $C_{\varepsilon,k} = B_k\,C_{\delta s}\,B_k^T + \Sigma_n$ with $B_k = \text{diag}(H_\xi, H_\eta)\,S_k$ and performing the block multiplication gives:
\begin{equation}
M = \frac{1}{4}\begin{bmatrix}H_\xi \\ H_\eta\end{bmatrix} C_{\delta s} \begin{bmatrix}H_\xi^T & H_\eta^T\end{bmatrix} + \bar{\sigma}^2 I
\label{eq:precond}
\end{equation}

This factored form is suitable for the Woodbury identity, leading to:
\vspace{-1em}
\begin{multline}
    M^{-1} = \frac{1}{\bar{\sigma}^2} I - \frac{1}{4\bar{\sigma}^4} \begin{bmatrix}H_\xi \\ H_\eta\end{bmatrix} \\
    \left(C_{\delta s}^{-1} + \frac{1}{4\bar{\sigma}^2} (H_\xi^T H_\xi + H_\eta^T H_\eta)\right)^{-1} \begin{bmatrix}H_\xi^T & H_\eta^T\end{bmatrix}
\end{multline}

The unitarity assumption ensures that the inner matrix simplifies to 
\begin{align}
    \left(C_{\delta s}^{-1} + \frac{1}{4 \bar{\sigma}^2}(H_\xi^T H_\xi + H_\eta^T H_\eta)\right)^{-1}
    &= \left(C_{\delta s}^{-1} + \frac{1}{2\bar{\sigma}^2}I\right)^{-1} \\
    &= F^T \Delta\, F
\end{align}
where $\Delta = \text{diag}(\Delta_j)$ acts as a spectral filter:
\begin{equation}
\Delta_j = \left(\frac{1}{\lambda_j} +\frac{1}{2\bar{\sigma}^2}\right)^{-1} = \frac{2\bar{\sigma}^2\lambda_j}{\lambda_j + 2\bar{\sigma}^2}
\label{eq:fourier_preconditionner}
\end{equation}
leading to:
\begin{equation}
M^{-1} = \frac{1}{\bar{\sigma}^2}I - \frac{1}{4\bar{\sigma}^4}\begin{bmatrix}H_\xi \\ H_\eta\end{bmatrix}(F^T\Delta\, F)\begin{bmatrix}H_\xi^T & H_\eta^T\end{bmatrix}
\label{eq:precond_inv}
\end{equation}

At low frequencies (speckle-dominated, $\lambda_j \gg 2\bar{\sigma}^2$), $\Delta_j \approx 2\bar{\sigma}^2$ whitens the correlated noise; at high frequencies (detector-dominated, $\lambda_j \ll 2\bar{\sigma}^2$), $\Delta_j \approx \lambda_j$ preserves spectral content. 

Applying $M^{-1}$ costs $\mathcal{O}(N^2 \log N)$ via FFT. The resulting speedup is quantified in Section~\ref{subsec:pcg_convergence}.

\subsection{PCG Convergence}
\label{subsec:pcg_convergence}

\begin{figure}
\begin{center}
\includegraphics[width=0.85\columnwidth]{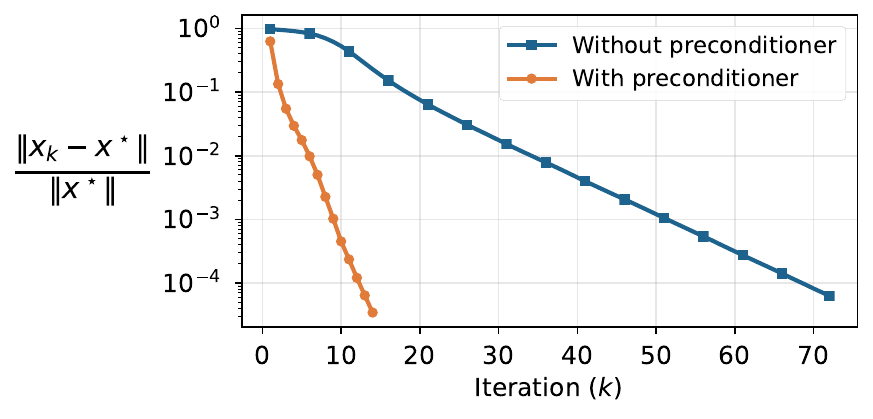}
\end{center}
\vspace{-8pt}
\caption{Convergence of the Conjugate Gradient solver with and without the Fourier-domain preconditioner.}
\label{fig:convergence}
\end{figure}

The preconditioner of Section~\ref{sec:preconditioner} dramatically accelerates convergence, as illustrated in Fig.~\ref{fig:convergence}. We plot the relative error $\nicefrac{\|x_k - x^\star\|_2}{\|x^\star\|_2}$ against the iteration number $k$, where $x^\star = C^{-1}b$ denotes the exact solution of the linear system $Cx = b$. The solver stops when the relative residual satisfies $\nicefrac{\|r_k\|_2}{\|b\|_2} \leq 10^{-5}$, with $r_k = b - Cx_k$. The preconditioned CG reaches this tolerance in 13 iterations versus 72 without preconditioning. Since this linear solve is called at every Score-ALD gradient step, this reduction translates directly into overall wall-clock speedup.

\section{Diffusion Prior and Conditional Sampling}
\label{sec:diffusion}

\subsection{Diffusion Process}

The prior on disk images $x=(I^u_\text{disk}, I^p_\text{disk})$ is 
learned via a DDPM \cite{ho_denoising_2020}. The forward process 
corrupts data with Gaussian noise:
\begin{equation}
x_t = \sqrt{\bar{\alpha}_t}\,x_0 + \sqrt{1-\bar{\alpha}_t}\,\epsilon, 
\quad \epsilon \sim \mathcal{N}(0, I)
\label{eq:diffusion_step}
\end{equation}
A network $\epsilon_\theta(x_t, t)$ is trained to predict $\epsilon$ 
by minimizing $\mathcal{L}_\text{simple}$ \cite{ho_denoising_2020}, 
which is equivalent to learning the score 
$\nabla_{x_t} \log p_t(x_t) \simeq 
-\epsilon_\theta(x_t,t)/\sqrt{1-\bar{\alpha}_t}$ 
\cite{vincent_connection_2011, song_score-based_2021}.

\subsection{Implementation Details}

The prior network $\epsilon_\theta$ is a U-Net with 80.4M parameters, trained for $\approx$40h on an NVIDIA A100 on 20\,000 synthetic $128\times128$ disk images generated with DDiT \cite{olofsson_challenge_2020}. The dataset covers a wide range of physical parameters, including semi-major axis, inclination, eccentricity, and opening angle; the polarization angle $\theta$ follows the orthoradial assumption \cite{engler_detection_2018} and is not a learned parameter. 
Training is performed exclusively on pristine disk images without stellar contamination, ensuring a clean separation between the learned morphological prior and the physical observation model.

\section{Experiments}

\subsection{Experimental Setup and Baselines}

\begin{figure}
\begin{center}
\includegraphics[width=0.7\columnwidth]{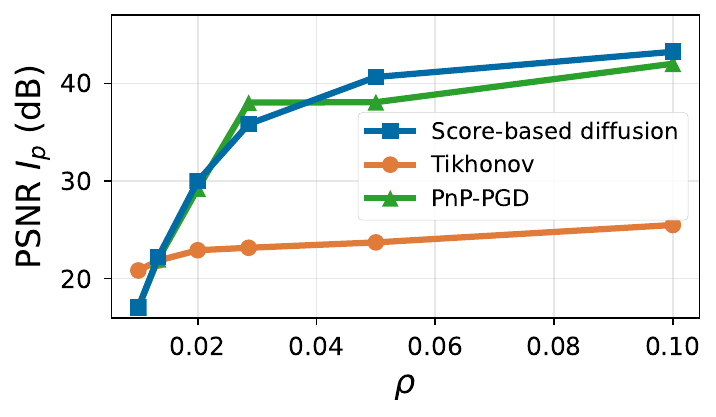}
\end{center}
\vspace{-8pt}
\caption{Reconstruction PSNR as a function of the disk-to-stellar 
peak intensity ratio $\rho$. Diffusion (blue), PnP 
\cite{hurault2022proximaldenoiserconvergentplugandplay} (green), 
and Tikhonov (orange); all baselines oracle-tuned.}
\label{fig:psnr_results}
\end{figure}

We evaluate reconstruction quality on synthetic data with speckle parameters $P_0 = 5000$, $\sigma_{\text{corr}} = 1.5$, across several stellar contamination levels defined by $\rho = \nicefrac{||x||_\infty}{||\bar{s}||_\infty}$, the ratio between the disk peak total intensity and the maximum stellar residual.

We also compare against a Plug-and-Play Proximal Gradient Descent (PnP-PGD) baseline \cite{hurault2022proximaldenoiserconvergentplugandplay} using the same trained U-Net to ensure a fair comparison. The PnP scheme iterates $x_{n+1} = D_{t_{\text{pnp}}}(x_n - \tau \nabla_x \mathcal{L}(x_n))$, where $\tau$ is the gradient step size and $\nabla_x \mathcal{L}$ is the likelihood gradient \eqref{eq:full_gradient} computed via PCG. Here, the denoiser 
$D_{t_{\text{pnp}}}(x_t) = \left( x_t - \sqrt{1-\bar{\alpha}_{t_{\text{pnp}}}}\,\epsilon_\theta(x_t, t_{\text{pnp}}) \right) / \sqrt{\bar{\alpha}_{t_{\text{pnp}}}} $ 
is the DDPM posterior mean estimator, used to directly predict the clean image $x_0$ at a fixed diffusion timestep $t_{\text{pnp}}$. This expression derives from~\eqref{eq:diffusion_step}.

The hyperparameters $t_{\text{pnp}} = 200$ and $\tau = 70$ were selected by grid search on a separate validation disk at $\rho = \nicefrac{1}{35}$ and kept fixed, running each reconstruction until convergence ($\approx 30$ steps). Unrolled approaches \cite{chappon_linear_2024} remain impractical here, as backpropagating through the PCG solver at each iteration would be prohibitively expensive.

\subsection{Results and Discussion}

As shown in Fig.~\ref{fig:psnr_results}, the blue curve shows the mean PSNR over $100$ different reconstructions from the score-based diffusion model (Algorithm~\ref{alg:score_ald_compact}) for the same disk. The orange and green curves correspond to Tikhonov and PnP baselines respectively; the Tikhonov regularization parameter $\lambda$ was oracle-tuned at each contamination level, while PnP hyperparameters are fixed at the values found by grid search.

At moderate to high SNR, both learned methods operate in a fundamentally different regime from Tikhonov: at $\rho = \nicefrac{1}{10}$, diffusion and PnP reach $43.2$ and $42.0$~dB respectively, versus $25.5$~dB for Tikhonov, a gap that highlights the benefit of learned priors over quadratic regularization. Within this regime, diffusion consistently outperforms PnP, as the generative prior more faithfully captures disk morphology. At the PnP tuning point ($\rho = \nicefrac{1}{35}$), both methods perform comparably ($30.0$ vs $29.2$~dB), with the fixed PnP hyperparameters precisely calibrated for this noise level. At very low SNR ($\rho = \nicefrac{1}{100}$), all three methods converge to similar performance ($17.0$~dB), where the observation is too corrupted for any prior to guide reconstruction reliably.

From a computational standpoint, PnP converges in $\sim$30 iterations versus $1000$ for Score-ALD --- a $30\times$ reduction, since both methods share the same per-step cost (one gradient and one network evaluation).

\begin{figure}
\begin{center}
\includegraphics[width=\columnwidth]{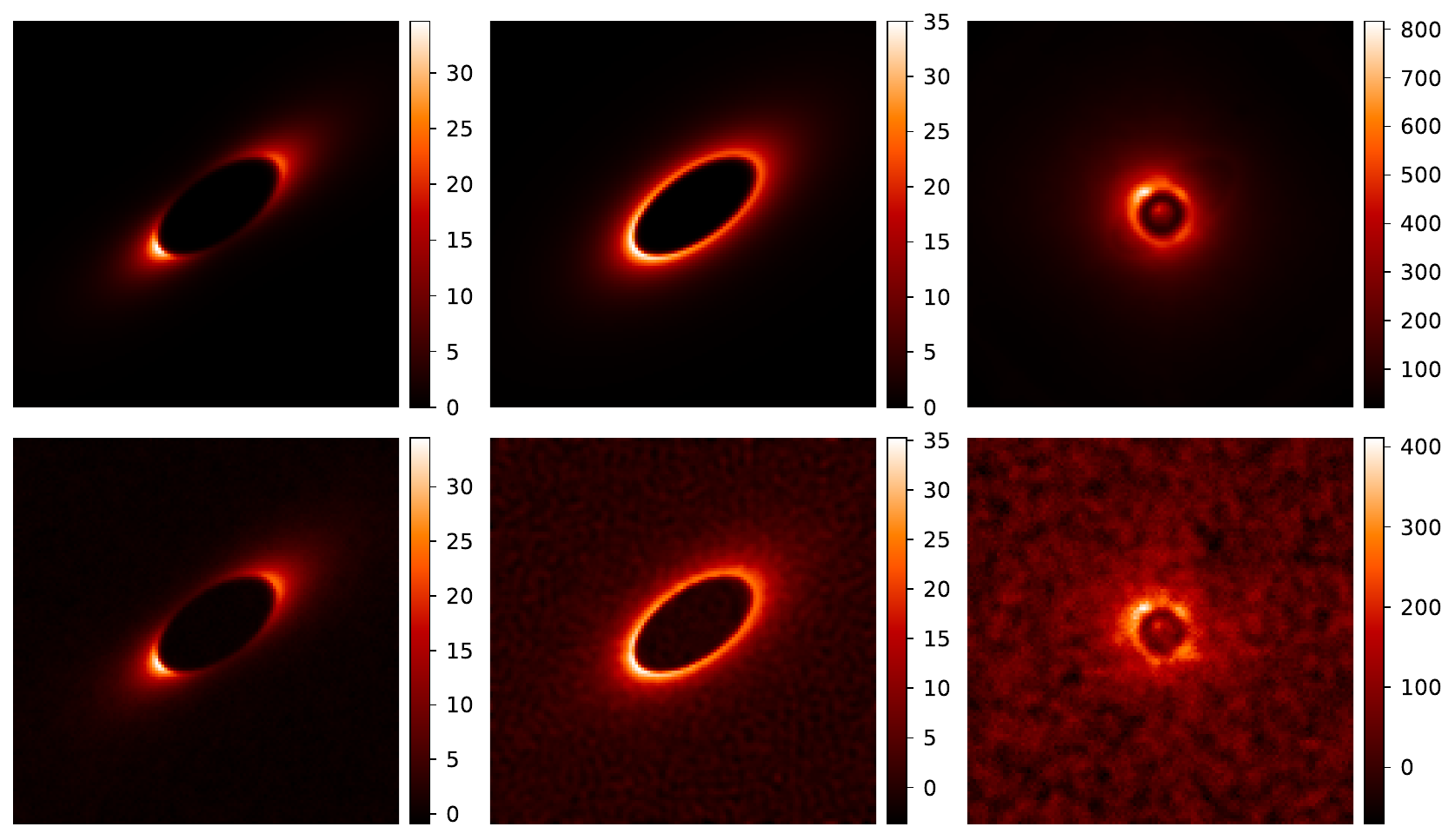}
\end{center}
\vspace{-8pt}
\caption{Top row: $I^p_\text{disk}$, $I_{\text{disk}}$ and $I_{\text{disk}}+I_{\text{star}}$. Bottom row: reconstructed $I^p_\text{disk}$, $I_{\text{disk}}$ and one of the detector measurements \eqref{eq:extended_model} (here $y_{0,1}$). \\ 
PSD parameters: $P_0=5000$ and $\sigma_{\text{corr}}=1.5$ \label{fig:recon}}
\end{figure}

Fig.~\ref{fig:recon} illustrates a representative reconstruction. The method successfully recovers both the polarized intensity $I^p_\text{disk}$ and the total disk intensity $I_\text{disk} = I^u_\text{disk}+I^p_\text{disk}$ from measurements dominated by stellar contamination. The ring morphology and fine structures are faithfully preserved in the reconstructed maps despite the raw measurement (bottom right of Fig.~\ref{fig:recon}) being visually dominated by stellar residuals.

The Gaussian model for speckle fluctuations, while tractable, allows unphysical negative values; in practice, for $P_0=5000$, $\sigma_{\text{corr}}=1.5$, around $3.3\%$ of simulated pixels are negative, confined to image edges far from the central region of interest, with negligible impact on reconstruction quality.
Furthermore, the method requires as inputs the mean speckle pattern $\bar{s}$; in a real observation pipeline, these quantities would need to be estimated from the data themselves or from auxiliary calibration frames. The diffusion model was trained and evaluated exclusively on synthetic DDiT data \cite{olofsson_challenge_2020}, and its generalization to real SPHERE/IRDIS observations remains to be demonstrated. Finally, despite the $5\times$ speedup provided by the preconditioner, the PCG solve at every Score-ALD step remains the computational bottleneck: a full 1000-step reconstruction currently takes approximately 20min on NVIDIA Quadro RTX 6000.

\section{Conclusion}

We presented a diffusion-based framework for polarimetric reconstruction of circumstellar disks that explicitly accounts for correlated speckle noise. The main contributions are: (i) an extended forward model separating disk convolution from speckle contamination with a PSD-parameterized covariance, (ii) a Fourier-domain preconditioner enabling a $5\times$ acceleration of the conjugate gradient solver, and (iii) a disk-only diffusion prior combined with Score-ALD sampling. The method outperforms optimally-tuned Tikhonov regularization and performs on par with the PnP baseline 
sharing the same U-Net backbone, with diffusion leading at high SNR 
and PnP at its tuning point ($\rho = \nicefrac{1}{35}$).

Future work will address non-stationary speckle models, faster diffusion sampling strategies \cite{daras_survey_2024}, uncertainty quantification via posterior variance, adaptive preconditioning, joint Angular / Polarimetric Differential Imaging processing, and validation on real SPHERE/IRDIS data.

\section*{Acknowledgment}
This work benefited from two French State grants  managed by the Agence Nationale de la Recherche (ANR) under the France 2030 program: the PEPR ORIGINS for the AIDA targeted project under reference ``ANR-22-EXOR-0016'', and the ANR DDISK project under reference ``ANR-21-CE31-0015''. 

This work was also granted access to the AI computing and storage resources at IDRIS through the resource allocation AD010415910 awarded by GENCI on the V100/A100 partition of the Jean Zay supercomputer.

\bibliographystyle{IEEEtran}
\bibliography{bibliography/references}


\end{document}